\documentclass[a4paper,twoside,12pt]{article}

\usepackage{psfig}
\usepackage{amsfonts}
\usepackage{amssymb}

\newcommand{\ZZ}{\mathbb{Z}}

\title{On a core instability of 't Hooft Polyakov monopoles}
\author{F.A.Bais$^1$ and J.Striet$^2$\\[2mm]
Institute for Theoretical Physics \\
University of Amsterdam \\
Valckenierstraat 65 \\
1018XE Amsterdam \\
The Netherlands\\
\date{May 2002}}

\begin{document}
\maketitle
\footnotetext[1]{bais@science.uva.nl}
\footnotetext[2]{jelpers@science.uva.nl}
%\tableofcontents

\begin{abstract}
\noindent We discuss a core instability of 't Hooft Polyakov monopoles in Alice
electrodynamics type of models in which charge conjugation symmetry is
gauged.  The monopole may deform into a toroidal defect which carries
an Alice flux and a (non-localizable) magnetic Cheshire charge.
\end{abstract}

\section{Introduction}
Since the pioneering work of 't Hooft and Polyakov
\cite{'thooft} magnetic monopoles have been studied in detail
in many different models. In this paper we address the question of
stability of the core of the fundamental, spherically symmetric,
monopole configuration, a stability which appears to be so obvious
that it was never seriously questioned. We will show that in a rather
simple model the spherically symmetric unit charge magnetic monopole
is not the global minimal energy solution for all parameter values in
the model. The fact that the core topology is not uniquely determined
by the boundary conditions and different core topologies can be
deformed into each other was already established earlier
\cite{bais2}. As we will indicate, Alice theories have a special
feature which makes it more plausible that such a core deformation
really may be favored energetically. Our interest in this problem was
rekindled by some observations that were made in theories with global
symmetries \cite{virga}.

We start by briefly summarizing the main features of Alice
Electrodynamics (AED), then we discuss the particular tensor model we
will use to explicitly establish the core instability and determine
some region in parameter space where this occurs.

\section{A core deformation in Alice Electrodynamics}
\begin{figure}[t,h,b]
\psfig{figure=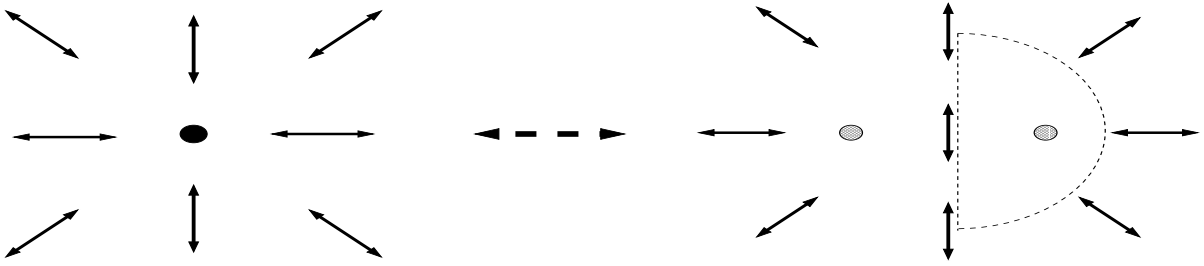,angle=0,width=16cm,height=3.5cm}
\caption[somethingelse]
{ \footnotesize A slice of the core deformation of the spherical
monopole into an Alice loop, carrying a magnetic Cheshire charge.}
\label{coredeform.eps}
\end{figure}

Alice electrodynamics (AED) is a gauge theory with gauge group
$H=U(1)\ltimes\ZZ_2\sim O(2)$, i.e. a minimally non-abelian extension
of ordinary electrodynamics. The nontrivial $\ZZ_2$ transformation
reverses the direction of the electric and magnetic fields and the
sign of the charges. In other words, in Alice electrodynamics charge
conjugation symmetry is gauged. However, as this non-abelian extension
is discrete, it only affects electrodynamics through certain global
(topological) features, such as the appearance of Alice fluxes and
Cheshire charges \cite{schwarz,alford}. The possibility of a
non-localizable magnetic Cheshire charge will be of great importance
in our study of the core instability of the monopole. The topological
structure of $U(1)\ltimes\ZZ_2$ differs from that of $U(1)$ in a few
subtle points. AED allows topologically stable localized fluxes since
$\Pi_0(U(1)\ltimes\ZZ_2) = \ZZ_2$, the so called Alice fluxes. Note
that in this theory this flux is co\"existing with the unbroken $U(1)$
of electromagnetism and is therefore not an ordinary ``magnetic''
flux. Just as a $U(1)$ gauge theory, AED may contain magnetic
monopoles, which follows from the fact that $\Pi_1(U(1)\ltimes\ZZ_2) =
\ZZ$. We note however, that due to the fact that the $\ZZ_2$ and the
$U(1)$ part of the gauge group do not commute, magnetic charges of
opposite sign belong to the same topological sector. Alice phases can
be generated by spontaneously breaking $SU(2)$ (or $SO(3)$) to
$U(1)\ltimes\ZZ_2$, for example by choosing a Higgs field in a
5-dimensional representation of the gauge group. In that case the
topological defects, fluxes \cite{NO,jelper} and monopoles
\cite{'thooft}, will correspond to regular classical
solutions. It was pointed out long ago that there are interesting
issues concerning the core stability of magnetic monopoles. Fixing the
asymptotics of the Higgs field, the core (i.e. the zeros of the Higgs
field\footnote{In fact in AED the Higgs field does not even need to go
to zero for the ring type solution}) may have different topologies,
notably that of a ring rather than the conventional point.  These core
topologies can be smoothly deformed into each other and it is a
question of energetics what will be the lowest energy monopole state
\cite{bais2}. We return to this issue in this paper because the core
deformation would be accompanied by the rather unusual delocalized
version of (magnetic) charge, the so called Cheshire charge. In the
specific AED model we studied the Higgs field is a symmetric tensor,
whose vacuum expectation value may be depicted as a bidirectional
arrow. The head-tail symmetry of the order parameter reflects the
charge conjugation symmetry of the theory. In AED the spherical
monopole can be punctured and be deformed into an Alice loop, this
configuration is consistent with the order parameter because of the
the head-tail symmetry of the order parameter. In figure
\ref{coredeform.eps} we show a slice of this core deformation.  Note
that the order parameter on the right hand side of figure
\ref{coredeform.eps} only rotates over an angle $\pi$ when going
around a single flux. This is the hallmark for an Alice flux, i.e.
the core deformed spherical monopole is in fact an Alice loop carrying
a magnetic Cheshire charge.

\section{The tensor model of AED}
To be able to answer stability questions of the spherically symmetric
monopole configuration (or the Cheshire charged Alice loop) we consider
an explicit model. In the remainder we focus on the original tensor
Alice model \cite{shankar}. The action of this model is given by:
\begin{equation}
S = \int d^4 x\ \{\frac{1}{4}F^{a,\mu\nu}F^a_{\mu\nu} + \frac{1}{4}Tr(
D^\mu \Phi D_\mu
\Phi )  - V(\Phi)\},
\end{equation}
where the Higgs field $\Phi = \Phi^{ab}$ is a real, symmetric,
traceless $3\times3$ matrix, i.e. $\Phi$ is in the five dimensional
representation of $SO(3)$ and $D_\mu \Phi =
\partial_\mu\Phi -ie[A_\mu,\Phi]$, with $A_{\mu}=A^a_{\mu}T_a$, where
$T_a$ are the generators of $SO(3)$. The most general renormalizeable
potential is given by
\cite{georgi}:
\begin{equation}
V=-\frac{1}{2}\mu^2Tr\Phi^2 -\frac{1}{3}\gamma Tr\Phi^3 + \frac{1}{4}
\lambda (Tr\Phi^2)^2
\label{potential}
\end{equation}
with $\gamma>0$, since $(\Phi,\gamma)=(-\Phi,-\gamma)$.\\ For a
suitable range of the parameters in the potential, the gauge symmetry
of the model will be broken to the symmetry of AED. In the ``unitary''
gauge, where the Higgs field is diagonal, the ground state is (up to
permutations) given by the following matrix:
\begin{equation}
\Phi_0=\left(\begin{array}{ccc}
-f&0&0\\ 0&-f&0\\ 0&0&2f
\end{array}\right)
\end{equation}
with $f =
\frac{\gamma}{12\lambda}(1+\sqrt{1+\frac{24\mu^2\lambda}{\gamma^2}})$.
The full action has four parameters, $e,\mu^2,\gamma,\lambda$, this
number can be reduced to two dimensionless parameters by appropriate
rescalings of the variables.  A physical choice for these
dimensionless parameters is to take the ratio's of the masses that one
finds from perturbing around the homogeneous minimum. To determine
these, we write the action in the unitary gauge where the massless
components of $\Phi$ have been absorbed by the gauge fields. The
physical components of the Higgs field may be expanded as:
\begin{equation}
\Phi(x^\mu)=\Phi_0+\sqrt{2}\phi_1(x^\mu) E_1
+\sqrt{2}\phi_2(x^\mu)~R_3(a(x^\mu))E_2R_3(a(x^\mu))^T
\end{equation}
with:
\begin{equation}
E_1=\frac{1}{\sqrt{6}}\left(\begin{array}{ccc} -1&0&0\\ 0&-1&0\\ 0&0&2
\end{array}\right)
~;~~E_2=\frac{1}{\sqrt{2}}\left(\begin{array}{ccc} 1&0&0\\ 0&-1&0\\
0&0&0
\end{array}\right)
~;~~E_3=\frac{1}{\sqrt{2}}\left(\begin{array}{ccc} 0&0&1\\ 0&0&0\\
1&0&0
\end{array}\right)
\end{equation}
and $R_i$ are the usual rotation matrices. To second order, the
potential $V(\Phi)$ takes the following form\footnote{It is most
convenient to use $\phi_2$ for the combination $\phi_2e^{ia}$, since
these two Higgs modes, $\phi_2$ and $a$, combine to form one complex
charged field, from now on called $\phi_2$.}:
\begin{equation}
V(\Phi) = const. + (2\mu^2 + \gamma f)\phi_1^2 + 3\gamma f|\phi_2|^2 +
...
\end{equation}
yielding the two distinct masses of the Higgs modes. Next we look at
the 'kinetic' term, $\frac{1}{4}Tr(D^\mu \Phi D_\mu \Phi )$, of the
Higgs field. Inserting the previous expressions for the Higgs field,
we find:
\begin{equation}
\frac{1}{4}Tr(D^\mu \Phi D_\mu \Phi) = \frac{1}{2}(\partial_\mu\phi_1)^2 +
\frac{1}{2}|D_\mu^3 \phi_2|^2 +\frac{9}{2}e^2f^2[(A_\mu^1)^2 + (A_\mu^2)^2] + ...
\end{equation}
with: $D_\mu^3=\partial_\mu - i 2 e A^3_\mu$. The second term shows
that the $\phi_2$ component of the Higgs field carries a charge $2e$
with respect to the unbroken $U(1)$ component $A_\mu^3$ of the gauge
field. The first term describes the usual charge neutral Higgs
particle and the third term yields the mass of the charged gauge
fields. Thus the relevant lowest order action is given by:
\begin{eqnarray}
S = \int d^4x \{ \frac{1}{4} F_{\mu\nu}^a F^{a,\mu\nu} +
\frac{1}{2}(\partial_\mu\phi_1)^2 + \frac{1}{2}|D_\mu^3\phi_2|^2 -
\frac{1}{2}m_1^2\phi_1^2 \\ \nonumber
- \frac{1}{2}m_2^2|\phi_2|^2 -
\frac{1}{2}m_A^2[(A_\mu^1)^2+(A_\mu^2)^2]+ ...\}
\end{eqnarray}
with $m_1^2=4\mu^2 + 2\gamma f$, $m_2^2=6\gamma f$ and
$m_A^2=9e^2f^2$. Thus two degrees of freedom of the five dimensional
Higgs field are 'eaten' by the broken gauge fields, one degree of
freedom forms the real neutral scalar field and two degrees of freedom
form the complex (doubly charged) scalar field. To specify a point in
the parameter space of classical solutions we may, up to irrelevant
rescalings, use the dimensionless mass ratio's $\frac{m_1}{m_2}$ and
$\frac{m_A}{m_2}$.

\section{The core instability}
\begin{figure}[t,h,b]
\psfig{figure=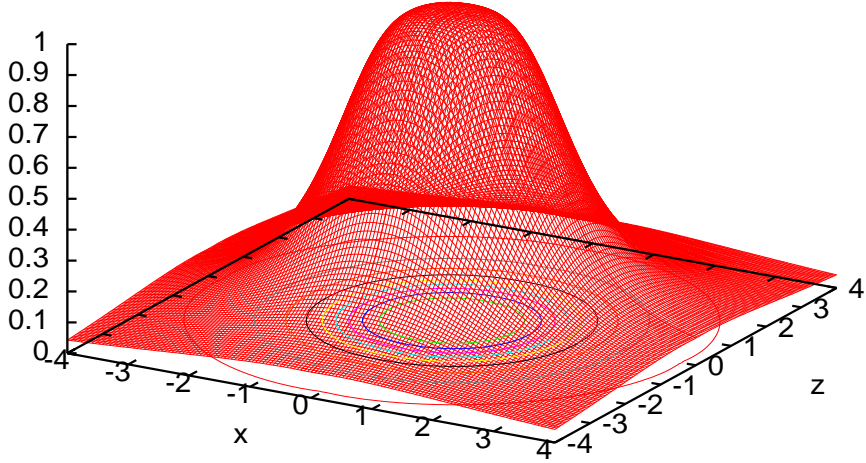,angle=270,width=14cm,height=7.5cm}
\caption[somethingelse]
{ \footnotesize A slice of a magnetic monopole solution at $y=0$, we
plotted $1-\frac{Tr\Phi^2}{f^2}$.\\ $~~~~~~$$\frac{m_1}{m_2}=0.571$
and $\frac{m_A}{m_2}=0.0095$.}
\label{monopole.eps}
\end{figure}

\begin{figure}[t,h,b]
\psfig{figure=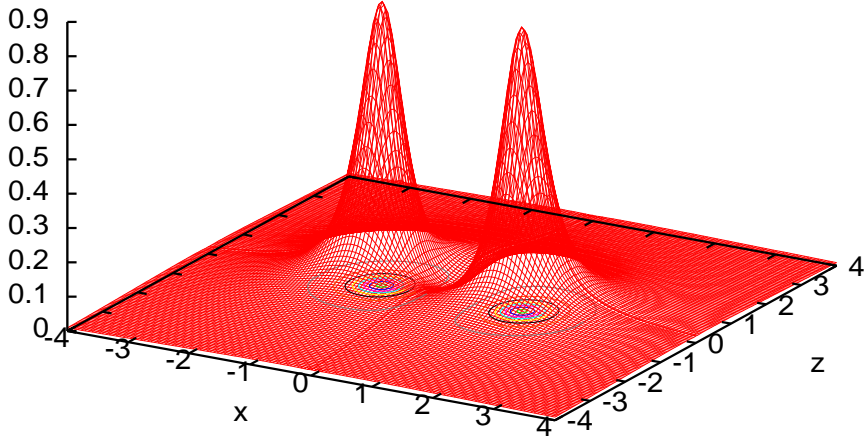,angle=270,width=14cm,height=7.5cm}
\caption[somethingelse] {
\footnotesize A slice of a Cheshire charged Alice loop configuration  at $y=0$, we
plotted $1-\frac{Tr\Phi^2}{f^2}$.\\$~~~~~~$ $\frac{m_1}{m_2}=0.882$
and $\frac{m_A}{m_2}=0.0073$.}
\label{aliceloop.eps}
\end{figure}

In this section we will show that the monopole core, see figure
\ref{monopole.eps}, becomes meta- or unstable for a certain range in
the parameter space of the theory. Our strategy is as follows. Using
numerical methods, we look for the global and local minima of the
monopole energy within a class of configurations given by a suitable
ansatz. We restrict ourselves to static configurations and the ansatz
we use contains the spherically symmetric 't Hooft-Polyakov monopole
configuration as a special case
\cite{shankar}. The ansatz is cylindrically symmetric and also has
reflection symmetry with respect to the $z=0$ plane.\\
\noindent The ansatz for the Higgs field is:
\begin{equation}
\Phi(z,\rho,\theta=0)=\phi_1(z,\rho) E_1+ \phi_2(z,\rho) E_2 + \phi_3(z,\rho) E_3
\end{equation}
and
\begin{equation}
\Phi(z,\rho,\theta)=R_3(\theta)\Phi(z,\rho,\theta=0)R_3(\theta)^T
\end{equation}
The ansatz for the gauge fields is simply given by
$eA_i^j=-\epsilon_{ijk}\frac{x^k}{x^2}A(z,\rho)$, very similar to the
one for the spherically symmetric monopole \cite{'thooft},
except that we allow $A(z,\rho)$ to depend on $\rho$ and $z$ instead
of only depending on $r=\sqrt{\rho^2+z^2}$. The boundary conditions for
$r\to\infty$ are the boundary conditions of the spherically symmetric
monopole as in \cite{shankar}, i.e. $A(z,\rho)$ goes to one and the
Higgs field to
$\Phi(z,\rho,\theta)=R_3(\theta)R_2(\arccos(\frac{z}{r}))
\Phi_0 R_2(\arccos(\frac{z}{r}))^TR_3(\theta)^T$.\\ The
boundary conditions for $\rho=0$ and $z=0$ follow by imposing the
cylindrical and reflection symmetry and are given in the table below:
\begin{displaymath}
\begin{array}{|c||c|c|}\hline
&\rho=0&z=0\\\hline
\phi_1&\partial_\rho\phi_1=0&\partial_z\phi_1=0\\
\phi_2&\partial_\rho\phi_2=\phi_2=0&\partial_z\phi_2=0\\
\phi_3&\phi_3=0&\phi_3=0\\
A&\partial_\rho A=0&\partial_zA=0\\\hline
\end{array}
\end{displaymath}
It is easy to see that these boundary conditions are also met by the
spherically symmetric monopole ansatz, so it is indeed contained in
our more general ansatz. The important point is that our ansatz in
principle allows for the possibility of an Alice loop configuration
carrying a magnetic Cheshire charge, see figure \ref{aliceloop.eps}.
These are exactly the two configurations that we want to compare.

\begin{figure}[t,h,b]
\psfig{figure=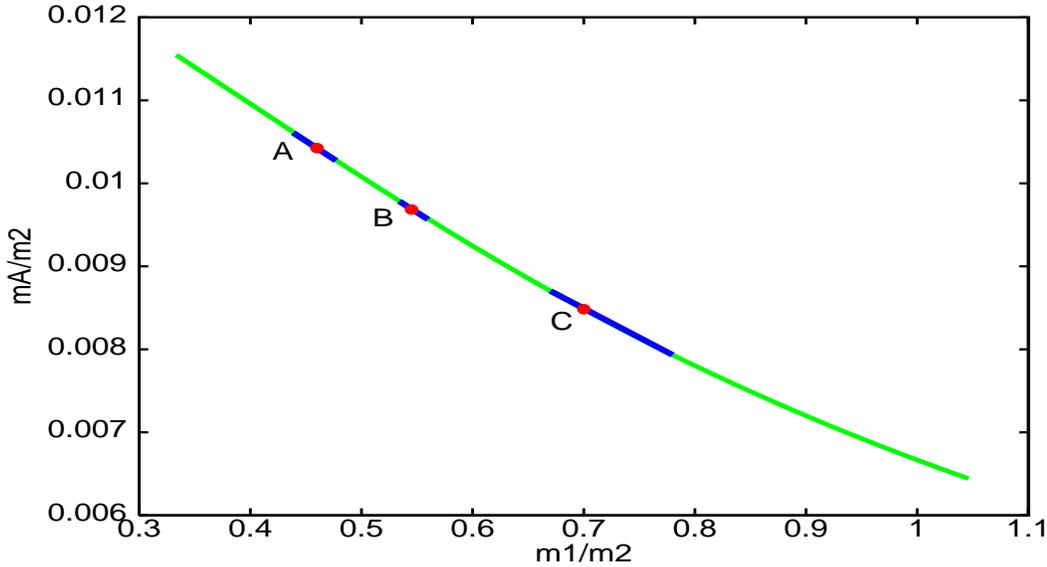,angle=270,width=14cm,height=7.5cm}
\caption[somethingelse]
{ \footnotesize Going from the left to the right: After point A, the
thick line represents the errorbar along the path, the Alice loop
becomes locally stable, between point A and C both the monopole and
the Alice loop are locally stable. After point B the monopole is only
meta stable and after point C it is even an unstable solution.}
\label{parampath.eps}
\end{figure}

Using the ansatz, we indeed found configurations having less energy
than the spherically symmetric monopole solution, at least in a
certain region of the parameter space. We even found that the
spherically symmetric monopole is not always locally stable. Strictly
speaking our non spherical symmetric configurations, the magnetically
Cheshire charged configurations, are only approximate solutions and
consequently, they only yield an upper bound to the energy of the true
solution. Obviously this suffices to show the instability of the
standard monopole and we do expect that the true solution is very
close to this magnetically Cheshire charged Alice loop
configuration. In this paper we only present the results concerning
configurations along a specific path in the parameter space of the
theory. We refer to an forthcoming paper \cite{jelper2} where we will
determine the stability, meta stability and instability regions of
both configurations, within this ansatz for the 'full' parameter space
of the model. The path, see figure
\ref{parampath.eps}, we have considered covers three regions of the
model. In one region, on the left of point A, the monopole is the only
stable solution. In the next region, between point A and C, both the
monopole and the Alice loop are locally stable and in the last, on the
right of point C, the Alice loop is the only locally stable
configuration.  Somewhere half way the middle region, point B, the
monopole is no longer the global minimum, whereas the Alice loop is,
i.e. the monopole is only a meta stable solution.\\
\\
The two extreme sides of the path can be understood as follows. The
two masses $m_1$ and $m_2$ correspond to the energy cost to deviate
from the vacuum in two different ways.  $m_1$ is the energy cost for
deviating in the neutral direction or 'length' of the Higgs field,
while $m_2$ is the energy cost in deviating in a non uniaxial
direction. Thus in the limit $\frac{m_1}{m_2}\to 0$, the deviations in
the non uniaxial directions are suppressed. There one would expect the
uniaxial monopole to be the global stable solution. In the opposite
limit, $\frac{m_1}{m_2}\to \infty$, one would expect an 'escape' into
the non-uniaxial directions and a suppression in the length deviation,
signaling the meta stability of the uniaxial monopole, as is the case
for the Alice loop configuration. Notice that the length of the Higgs
field never becomes zero\footnote{ Not shown here, but we also find
that the minimum length of the Higgs field in the case of the Alice
loop increases for increasing $\frac{m_1}{m_2}$ as this argument
indicates.} in the case of the Alice loop,
i.e. $1-\frac{Tr\Phi^2}{f^2}$ never becomes one in figure
\ref{aliceloop.eps}.

\section{Conclusion and outlook}
In this letter we showed that monopoles of the 't Hooft Polyakov type
may exhibit a core instability, depending on the parameters of the
theory. In one part of the parameter space, the spherical monopole is
the global minimum. In another part it corresponds to a local minimum
and there even is a region where it is unstable. We found that the
competing configuration is a magnetically Cheshire charged Alice loop.
Since we worked within a limited ansatz, the regions we found for
the monopole global and/or local stability are in fact only upper
bounds on the stability regions of the spherical monopole, and these
regions can only become smaller when no (or less) restrictions are put
on the configurations one may sample. At the moment we are scanning
the 'full' parameter space of the model. The results obtained as well
as more detailed information on the model, the simulations and the
configurations we found, will be published elsewhere \cite{jelper2}.

\end{document}